\documentclass[twocolumn, superscriptaddress]{revtex4}
\usepackage{amsmath}
\usepackage{graphicx}
\begin{document}
\title{Quantum Oscillations in the Chiral Magnetic Conductivity}
\author{Sahal Kaushik}
\email{sahal.kaushik@stonybrook.edu}
\affiliation{Department of Physics and Astronomy, Stony Brook University, Stony Brook, New York 11794, USA}
\author{Dmitri E. Kharzeev}
\email{dmitri.kharzeev@stonybrook.edu}
\affiliation{Department of Physics and Astronomy, Stony Brook University, Stony Brook, New York 11794, USA}
\affiliation{Department of Physics, Brookhaven National Laboratory, Upton, New York 11973-5000, USA}
\affiliation{RIKEN-BNL Research Center, Brookhaven National Laboratory, Upton, New York 11973-5000, USA}
\begin{abstract}
In strong magnetic field the longitudinal magnetoconductivity in 3D chiral materials is shown to exhibit a new type of quantum oscillations arising from the chiral magnetic effect (CME).  These quantum CME oscillations are predicted to dominate over the Shubnikov-de Haas (SdH) ones  in chiral materials with an approximately conserved chirality of quasiparticles at strong magnetic fields. The phase of quantum CME oscillations differs from the phase of the conventional  SdH oscillations by $\pi/2$.
\end{abstract}
\maketitle
The chiral magnetic effect (CME) \cite{Fukushima:2008xe} (see \cite{Kharzeev:2013ffa,Kharzeev:2012ph} for reviews and additional references) is a macroscopic quantum transport phenomenon induced by the chiral anomaly. In 3D systems possessing chiral fermions, an imbalance between the densities of left- and right-handed fermions generates a non-dissipative electric current along the direction of an external magnetic field. The CME has been predicted to occur in Dirac and Weyl semimetals (DSMs/WSMs) \cite{Fukushima:2008xe,Son:2012wh,Son:2012bg,Zyuzin:2012tv,Basar:2013iaa,vazifeh2013electromagnetic,goswami2013axionic}, and has been recently experimentally observed through the measurement of negative longitudinal magnetoresistance in DSMs \cite{Li:2014bha,kim2013dirac,xiong2015evidence,li2015giant} as well as in WSMs  \cite{huang2015observation,wang2015helicity,zhang2015observation,yang2015observation,shekhar2015large,yang2015chiral}. 

The CME electric current flowing along the  external magnetic field $\bf{B}$ in the presence of a chiral chemical potential $\mu_5$ is given by
\begin{equation}\label{eq1}
  {\bf{j}_\mathrm{CME}} = \frac{e^2}{2\pi^2}\ \mu_5\ \bf{B}.
\end{equation}
The chiral chemical potential $\mu_5 = (\mu_R - \mu_L)/2$ vanishes in equilibrium, but can be created by the chiral anomaly \cite{adler1969r,jackiw} once external parallel electric and magnetic fields are applied; the relevance of chiral anomaly in condensed matter systems was first pointed out in \cite{nielsen1983adler}.  Other possibilities include pumping the system by  light \cite{zhong2016gyrotropic}, strain \cite{cortijo2016strain}, pseudoscalar phonons \cite{song2016detecting}, or using the asymmetric Weyl semimetals driven by AC voltage  \cite{kharzeev2016kikuchi}. 

The rate of chirality production in parallel electric and magnetic fields due to the chiral anomaly  is given by  
\begin{equation}
\dot{\rho_5} = \frac{e^2}{2\pi^2}\ {\bf{E}}\cdot{\bf{B}} - \frac{\rho_5}{\tau_V}, 
\end{equation}
where $\rho_5 = \rho_R - \rho_L$ is the difference between the densities of the right-handed and left-handed fermions, and the second term is introduced to take account of the chirality changing transitions with a characteristic time $\tau_V$. If the chirality flipping time $\tau_V$ is much greater than the scattering time $\tau$, the left-handed fermions and the right-handed fermions can exist in a steady state with different chemical potentials $\mu_L$ and $\mu_R$. 

In a uniform and constant magnetic field, the energies of the lowest Landau levels are $\epsilon = -vp_z$ for left-handed and $\epsilon = + vp_z$ for right-handed fermions, see Fig. 1 ($v$ is the Fermi velocity; we assume that magnetic field $B$ is along the $z$-axis). The energies of excited Landau levels are $E = \pm v\sqrt{p_z^2 + 2eBn}$ for $n\geq 1$, for both chiralities.  The density of Landau levels in the $xy$ plane is $eB/2\pi$, whereas the density of states in the $z$-direction is $p_z/2\pi$.
Because the lowest Landau level is not degenerate in spin, it has right-handed fermions of positive charge traveling along $B$ and left-handed ones of negative charge traveling in the opposite direction. This induces the CME current (\ref{eq1}).

The density of the chiral charge $\rho_5$ is related to the chiral chemical potential $\mu_5$ through the chiral susceptibility $\chi \equiv \partial \rho_5  / \partial \mu_5$ -- at small $\mu_5$, $\rho_5 = \chi \mu_5 + ...$ so that $\mu_5 \simeq  \chi^{-1} \rho_5$. Note that in the absence of chirality loss corresponding to $\tau_V \to \infty$ the CME current would grow linearly in time -- in other words, it would behave as a superconducting current, see \cite{kharzeev2016chiral} for a discussion. 

At finite $\tau_V$, the density of the chiral charge saturates at the value $\rho_5 = e^2/2\pi^2\ {\bf{E}}\cdot{\bf{B}} \ \tau_V$, and the longitudinal CME conductivity for parallel ${\bf E}$ and ${\bf B}$ is given by 
\begin{equation}\label{CMEsigma}
\sigma_{\rm CME} = \frac{e^4 B^2}{4\pi^4 \chi (i\omega + 1/\tau_V)} ,
\end{equation}
where $\omega$ is the frequency of an external field.

\begin{figure}[ht]\label{landau}
\begin{center}
\includegraphics[scale=0.5]{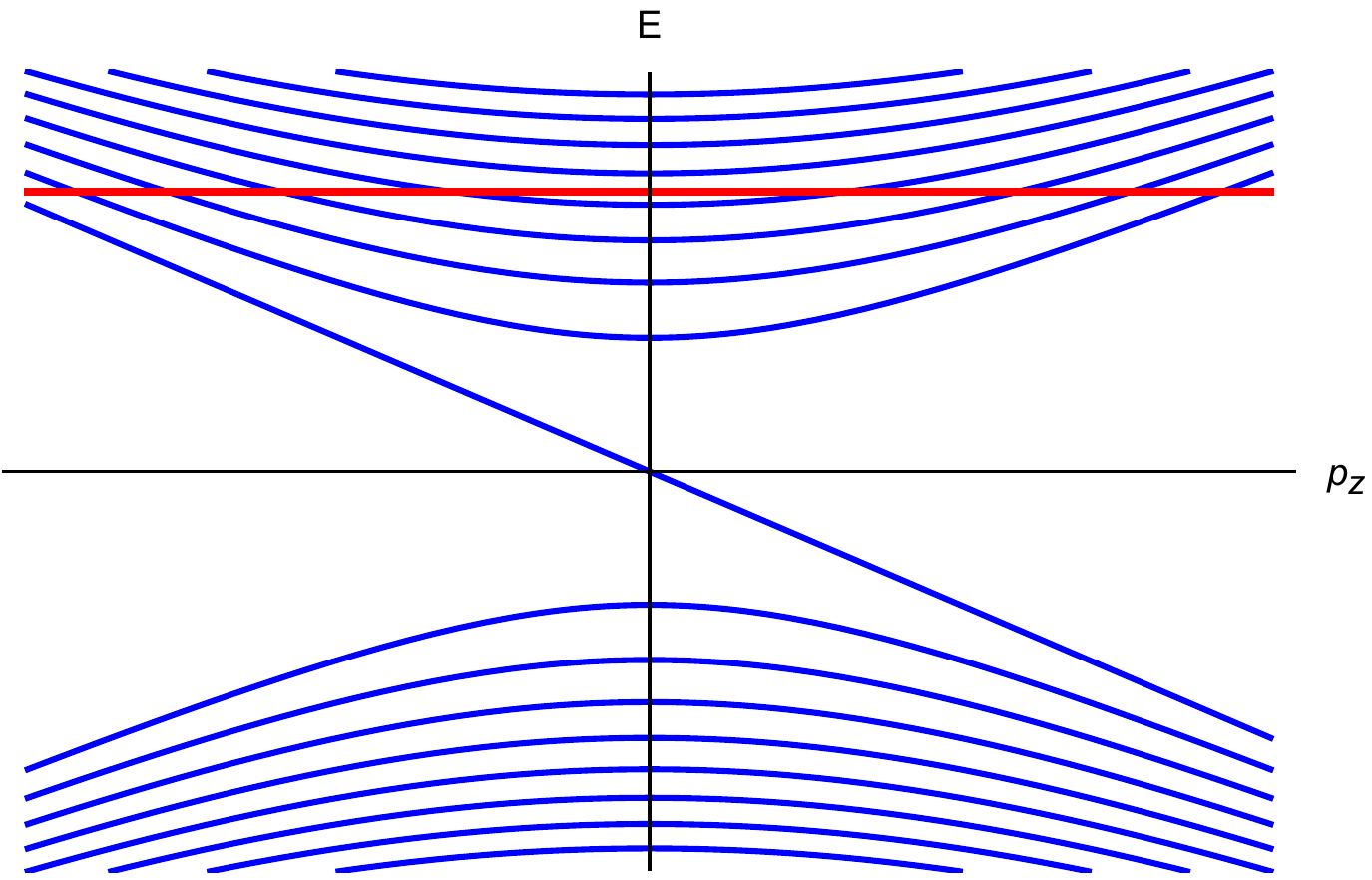} \\~\\ \includegraphics[scale=0.5]{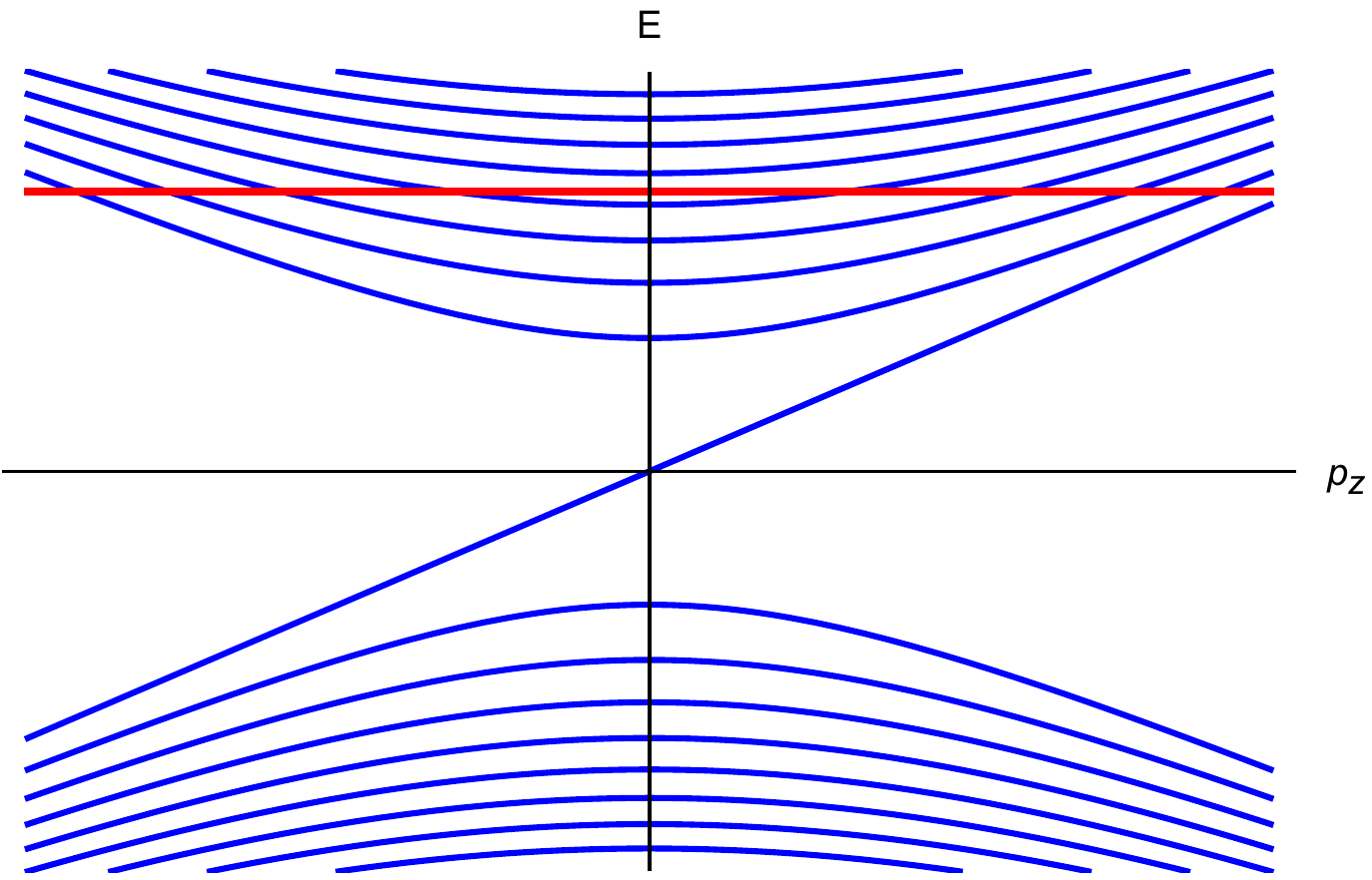}
\end{center}
\caption{The energy structure of left-handed and right-handed fermions in a magnetic field. The red line indicates $\mu$.}
\end{figure}

In the presence of a Fermi surface with a chemical potential $\mu = (\mu_R + \mu_L)/2$, in weak magnetic fields with $2eBv^2 \ll \mu^2$ the Landau quantization can be ignored, and the chiral susceptibility is given by
\begin{equation}\label{susc_weak}
\chi = \frac{\mu^2}{\pi^2 v^3} + \frac{T^2}{3v^3}.
\end{equation}
The DC CME conductivity \cite{Fukushima:2008xe,Son:2012bg, Zyuzin:2012tv, Li:2014bha} is then
\begin{equation}
\sigma_{\rm CME} = \frac{e^4 v^3 \tau_V B^2}{4\pi^2 (\mu^2 + \pi^2 T^2/3)} .
\end{equation}
In strong magnetic fields with $2eBv^2 \gg \mu^2$, only the lowest Landau level contributes, and $\mu = eB/2v\pi^2$, so  the DC CME conductivity has a linear dependence on $B$: 
\begin{equation}
\sigma_{\rm CME} = \frac{e^2 v\ \tau_V B}{2\pi^2}.
\end{equation}
 In a real material, the conductivity is the sum of the CME conductivity and the Ohmic conductivity,  $\sigma_{zz} = \sigma_{\rm CME} + \sigma_{\rm Ohm}$. 

As is well known, the Shubnikov - de Haas (SdH) oscillations appear 
due to the quantized Landau levels and the presence of Fermi surface; they  have been observed in the transverse magnetoconductivity of some Dirac and Weyl (semi)metals \cite{liang2015ultrahigh, huang2015observation, hu2016pi}. The phase of the SdH oscillations depends on the Berry curvature; this fact can be used to distinguish materials with massive carriers from Dirac and Weyl materials \cite{huang2015observation, hu2016pi, murakawa2013detection, luk2006dirac}. The oscillating part of the transverse conductivity has the form
\begin{equation}
\sigma_{xx} = A(B) \cos\left[2\pi\left(\frac{B_0}{B} - \gamma + \delta\right)\right] ,
\end{equation}
where $\gamma$ is $0$ for Dirac and Weyl carriers and $1/2$ for massive carriers, and $\delta$ varies between $-1/8$ and $1/8$ for 3D materials.

In this work, we point out the existence of a new type of quantum oscillations that emerge in strong magnetic fields due to a non-linear relation between the chiral chemical potential $\mu_5$ and the density of chiral charge $\rho_5$. 
Our treatment will apply to Dirac materials and Weyl materials in which the Weyl points have the same energy and Fermi velocity. The assumptions used this work are $\tau_V \gg \tau$ (i.e. the chirality flips are relatively rare), $T\ll\mu$, $\mu_5\ll T, \mu$ and 
$2eBv^2 \ll \mu^2$; we set $\hbar = 1$.


The density of states in energy $E$ for each chirality is
\begin{equation}\label{DoS}
g(E) = \frac{E_L^2}{8\pi^2 v^3}\left[1+2\sum_{n=1}^\infty \Theta(E^2 - nE_L^2) \sqrt{\frac{E^2}{E^2 - n E_L^2}}\right]
\end{equation}
where $E_L^2 = 2eBv^2$ is the difference in the squares of the Landau level energies. The factor of $2$ is because we have particles traveling in both directions for higher levels.

The total number density of particles, for each chirality is given by
\begin{equation}
\rho_{R,L} (\mu, T) = \int_{E_-}^{E_+} g(E) f(\mu - E, T) dE ,
\end{equation}
where $f(V, T) = \frac{e^{V/T}}{e^{V/T}+1}$ is the Fermi distribution function, and $E_-, E_+$ are the cutoff energies. Therefore,
\begin{equation}
\chi = \frac{\partial \rho_5}{\partial \mu_5} = 2 \frac{\partial \rho_{L,R}}{\partial \mu_{R,L}} = 2 \int_{E_-}^{E_+} g(E) f'(\mu - E, T) dE
\end{equation}
yielding
\begin{align}
\chi = \frac{E_L^2}{2\pi^2 v^3}  \int_{-\infty}^\infty & \left[\frac{1}{2}+\sum_{n=1}^\infty \Theta(E^2 - nE_L^2) \sqrt{\frac{E^2}{E^2 - n E_L^2}}\right] \nonumber\\ &\times f'(\mu-E,T) dE
\end{align}
For small fields (when many Landau levels contribute), the sum can be approximated by an integral, and we recover (\ref{susc_weak}):
\begin{equation}
\chi(B=0) = \frac{\mu^2}{\pi^2 v^3} + \frac{T^2}{3v^3} .
\end{equation}
Let us now evaluate the quantum corrections to this expression that will be responsible for the quantum oscillations in CME conductivity. We are concerned only with energies close to $\mu$, so 
\begin{align}
\sqrt{\frac{E^2}{E^2 - n E_L^2}} &\approx \sqrt{\frac{\mu^2}{\mu^2 + 2\mu V - n E_L^2}} \\ &= \sqrt{\frac{\mu}{2(V-(n-x)(E_L^2/2\mu))}} ,
\end{align}
where $x = \mu^2/E_L^2$ and $V = E-\mu$.

We can define the contribution of the $n$th level to the susceptibility $\chi_n$ as
\begin{align}
\chi_n = \frac{E_L^2}{2\pi^2 v^3}\int_{-\infty}^\infty &\Theta(E^2 - nE_L^2) \sqrt{\frac{E^2}{E^2 - n E_L^2}}f'(\mu-E,T) dE \\ \approx \frac{E_L^2}{2\pi^2 v^3}\int_{-\infty}^\infty &\Theta(V-(n-x)(E_L^2/2\mu)) \nonumber \\ &\times \sqrt{\frac{\mu}{2(V-(n-x)(E_L^2/2\mu))}} f'(V,T) dV .
\end{align}
Now, in the expression
\begin{equation}
\chi = \sum_{n=1}^\infty \chi_n + \frac{1}{2} \chi_0
\end{equation}
we can extend the sum to $-\infty$ and approximate the contribution of the fictitious negative levels by an integral:
\begin{align}
\chi &\approx \sum_{n=-\infty}^\infty \chi_n - \int^0_{-\infty} \chi_n dn \\ &= \sum_{n=-\infty}^\infty \chi_n - \int^\infty_{-\infty} \chi_n dn + \int^\infty_0 \chi_n dn \\ &= \sum_{n=-\infty}^\infty \chi_n - \int^\infty_{-\infty} \chi_n dn + \frac{\mu^2}{\pi^2 v^3} + \frac{T^2}{3v^3}
\end{align}
We can then use the Poisson summation in 
\begin{equation}
\chi = \frac{\mu^2}{\pi^2 v^3} + \frac{T^2}{3v^3} + \sum_{l=1}^\infty 2\Re (\chi_l)
\end{equation}
to evaluate the Fourier transform of $\chi_n$,
\begin{equation}
\chi_n = \alpha\left[(x-n)\frac{E_L^2}{2\mu}\right]
\end{equation}
where 
\begin{align}
\alpha(z) &= \frac{E_L^2}{2\pi^2 v^3}\int_{-\infty}^\infty \Theta(V+z)\sqrt{\frac{\mu}{2(V + z))}} f'(V,T) dV \\ &\equiv \frac{E_L^2}{2\pi^2 v^3}(\beta\star\gamma)(z) ,
\end{align}
with $\beta(z) = \Theta(z)\sqrt{\frac{\mu}{2z}}$ and $\gamma(z) = f'(z,T)$. Here we have used the fact that $f'(V,T)$ is even in $V$. So according to the Poisson summation,
\begin{equation}
\chi_l = \frac{2\mu}{E_L^2}\sqrt{2\pi}\exp(-2\pi i l x) \tilde{\alpha}\left(2\pi l \frac{2\mu}{E_L^2}\right) .
\end{equation}
From the convolution theorem, 
\begin{equation}
\tilde{\alpha}(k) = \frac{E_L^2}{2\pi^2 v^3} \sqrt{2\pi} \tilde{\beta}(k)\tilde{\gamma}(k) .
\end{equation}
The Fourier transforms are
\begin{equation}
\tilde{\beta}(k) = \sqrt{\frac{\mu}{2}}\frac{1}{2}\frac{|k| + ik}{|k|^{\frac{3}{2}}} ,
\end{equation}
\begin{equation}
\tilde{\gamma}(k) = \sqrt{\frac{\pi}{2}}\frac{kT}{\sinh(\pi kT)} ,
\end{equation}
and
\begin{equation}
\chi_l = \frac{\mu E_L}{4\pi^2 v^3} \frac{(1+i)}{\sqrt{l}}\frac{\left(l\frac{4\pi^2 \mu T}{E_L^2}\right)}{\sinh\left(l\frac{4\pi^2 \mu T}{E_L^2}\right)} \exp(-2\pi i l \mu^2/E_L^2) .
\end{equation}
In a real material, the scattering caused by impurities smears the Landau levels. The density of states is thus the convolution of (\ref{DoS}) with a Lorentzian distribution $\frac{1}{\Gamma\pi}\frac{\Gamma^2}{\Gamma^2 + (E-E_0)^2}$, where $\Gamma$ is the Dingle factor. Therefore, we must multiply each harmonic in the oscillating term by factor of $\exp(-4\pi l\Gamma\mu/E_L^2)$ (the Fourier transform of the Lorentzian):
\begin{align}\label{chi_sus}
\chi \approx &\frac{\mu^2}{\pi^2 v^3} + \frac{T^2}{3v^3} + \frac{\mu E_L}{2\pi^2 v^3} \sum_{l=1}^\infty \frac{1}{\sqrt{l}}\frac{\left(l\frac{4\pi^2 \mu T}{E_L^2}\right)}{\sinh\left(l\frac{4\pi^2 \mu T}{E_L^2}\right)}\nonumber \\ &\times \exp(-4\pi l\Gamma\mu/E_L^2) [\cos(2\pi l \mu^2/E_L^2) + \sin(2 \pi l \mu^2/E_L^2)]
\end{align}
\begin{figure}[ht]
\begin{center}
\includegraphics[scale=0.65]{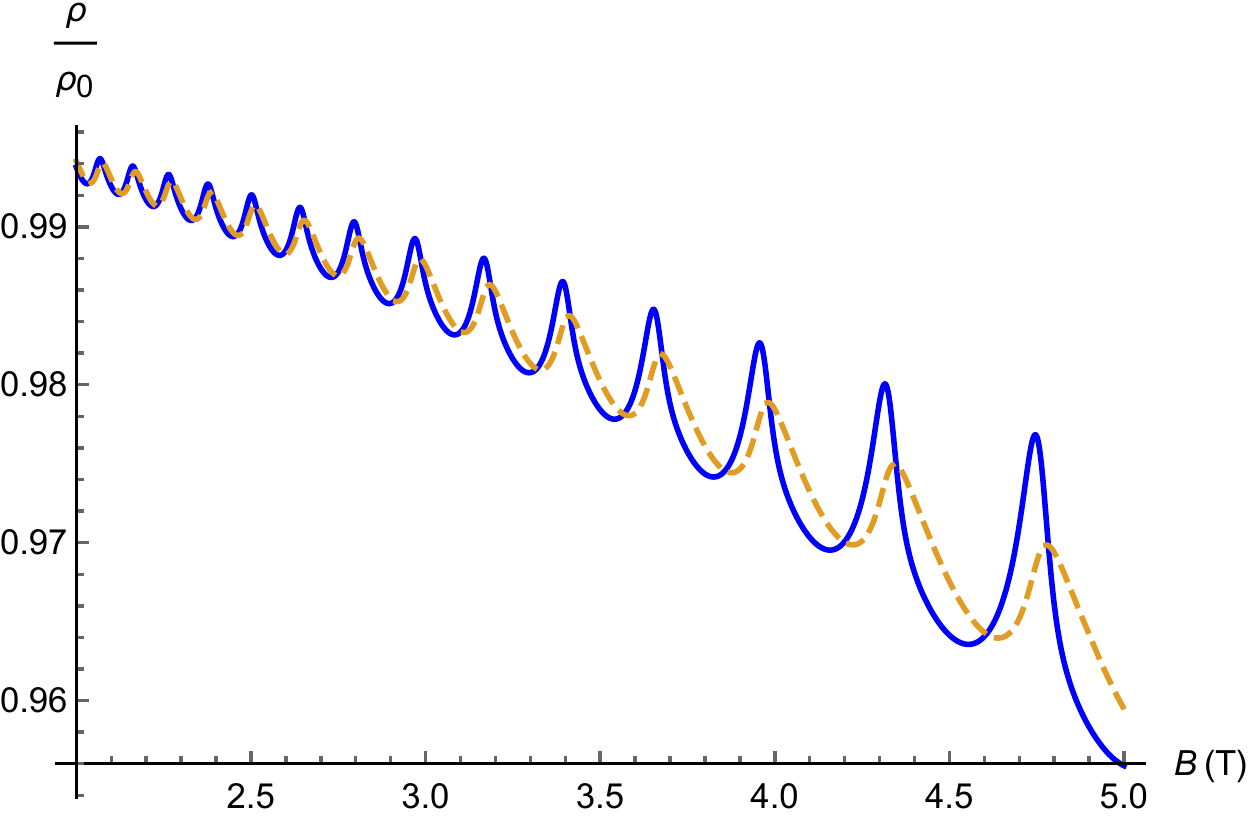}
\end{center}
\caption{$\rho_{zz}/\rho_0$ vs $B$ for $\mu = 150\ \mathrm{meV}$, $v = c/600$,  $T = 1.74\ \mathrm{K}$, $\Gamma = 0.3\ \mathrm{meV}$, and $\tau_V/\tau = 20$.  The solid line represents the full prediction taking account of the quantum CME oscillations, see (\ref{Total}); the dashed line represents only the SdH oscillations given by (\ref{SdH}). The quantum CME oscillations become larger than the SdH oscillations at $B \simeq 3$ T.}
\end{figure}
\begin{figure}[]
\begin{center}
\includegraphics[scale=0.65]{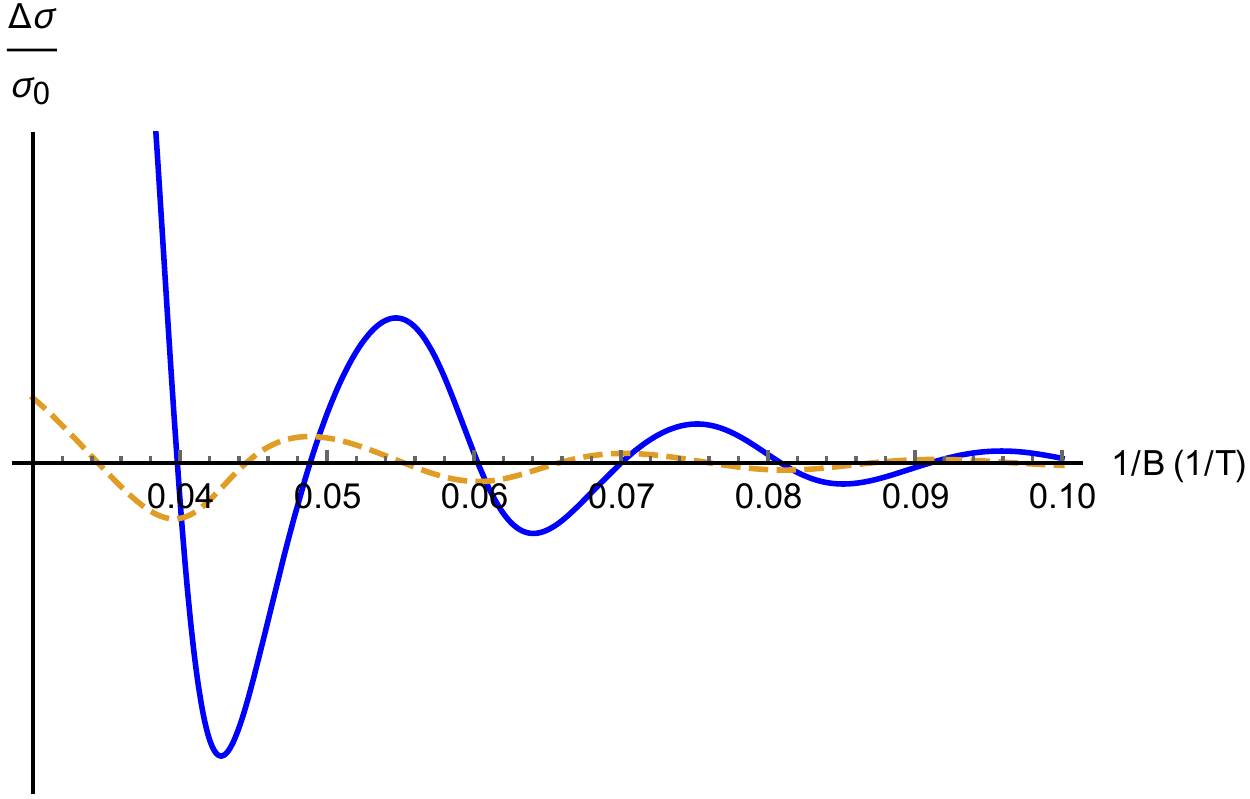}
\end{center}
\caption{The residue of $\sigma_{zz}$ after subtracting the constant and quadratic in $B$ contributions, plotted as a function of $1/B$ for $\mu = 150\ \mathrm{meV}$, $v = c/600$,  $T = 34.8\ \mathrm{K}$, $\Gamma = 0.3\ \mathrm{meV}$, and $\tau_V/\tau = 20$. 
The solid line represents the prediction of (\ref{Total}) while the dashed line represents the predictions of (\ref{SdH}) that ignore the quantum CME oscillations.}
\end{figure}

Since $\chi$ oscillates as a function of magnetic field, the CME conductivity also acquires these quantum oscillations. 
The Ohmic conductivity also oscillates with $B$; these oscillations for 3D chiral materials have been evaluated using the chiral kinetic theory in \cite{monteiro2015magnetotransport}. In our notations,
\begin{align}\label{SdH}
\frac{\sigma_{zz} (B)}{\sigma_0} \approx & 1 + \frac{3}{16}\frac{\tau_V}{\tau} \frac{E_L^4}{\mu^4} - \frac{3}{20} \frac{E_L^4}{\mu^4} - \frac{3}{8\pi} \frac{E_L^3}{\mu^3} \sum_{l=1}^\infty \frac{1}{l^{3/2}}  \nonumber\\ &\times \frac{\left(l\frac{4\pi^2 \mu T}{E_L^2}\right)}{\sinh\left(l\frac{4\pi^2 \mu T}{E_L^2}\right)}\exp(-4\pi l\Gamma\mu/E_L^2)\nonumber\\ &\times[\cos(2\pi l \mu^2/E_L^2) - \sin(2\pi l \mu^2/E_L^2)] ,
\end{align}
where $\tau$ is the (chirality-preserving) scattering time and $\sigma_0 \equiv \sigma(B=0) = \frac{\mu^2 e^2 \tau}{3\pi^2 v^2}$. The $\frac{3}{16}\frac{\tau_V}{\tau} \frac{E_L^4}{\mu^4}$ term, which is quadratic in $B$, comes from the CME conductivity;  to account for the variation of $\chi$ with $B$, we should now include a factor of $\frac{\mu^2}{\pi^2 v^3 \chi}$ in this term. Note that in weak magnetic fields, according to (\ref{susc_weak}), this factor is equal to unity, $\frac{\mu^2}{\pi^2 v^3 \chi} = 1$, but quantum corrections to $\chi$ given by (\ref{chi_sus}) will now induce additional oscillations in longitudinal magnetoconductivity. All other terms in (\ref{SdH}) represent the Ohmic conductivity. Therefore, the total longitudinal conductivity as a function of $E_L = \sqrt{2eB} v$ is
\begin{align}\label{Total}
\frac{\sigma_{zz} (B)}{\sigma_0} \approx & 1 + \frac{3}{16}\frac{\mu^2}{\pi^2 v^3 \chi}\frac{\tau_V}{\tau} \frac{E_L^4}{\mu^4} - \frac{3}{20} \frac{E_L^4}{\mu^4}  \nonumber \\ &- \frac{3}{8\pi} \frac{E_L^3}{\mu^3} \sum_{l=1}^\infty \frac{1}{l^{3/2}}  \frac{\left(l\frac{4\pi^2 \mu T}{E_L^2}\right)}{\sinh\left(l\frac{4\pi^2 \mu T}{E_L^2}\right)} \exp(-4\pi l\Gamma\mu/E_L^2)\nonumber \\ &\times [\cos(2\pi l \mu^2/E_L^2) - \sin(2\pi l \mu^2/E_L^2)] ,
\end{align}
where the chiral susceptibility $\chi$ that enters the second term oscillates with $B$ according to (\ref{chi_sus}).
When the temperature or the Dingle factor are large enough so that the first term in the Fourier series dominates, the longitudinal conductivity is given by
$$
\frac{\sigma_{zz} (B)}{\sigma_0} \approx  1 + \left(\frac{3}{16}\frac{\tau_V}{\tau} - \frac{3}{20}\right) \left(\frac{B}{B_0}\right)^2\nonumber\\ -
$$
\begin{equation}\label{approx}
- A(B) \left[\cos\left(\frac{B_0}{B}+\frac{\pi}{4}\right) + \frac{\pi}{4}\frac{\tau_V}{\tau}\frac{B}{B_0}\cos\left(\frac{B_0}{B}-\frac{\pi}{4}\right)\right] ,
\end{equation} 
where $B_0 = \mu^2/2ev^2$ and $A(B)$ is a positive non-oscillating factor which represents the effects of the temperature and the Dingle factor. For a material with $\mu = 150\ \mathrm{meV}$ and Fermi velocity $v = c/500$, the value of $B_0$ is $B_0 \approx 48\ \mathrm{T}$.

When chirality flipping time is much longer than the scattering time $\tau_V/\tau \gg 1$, and in strong magnetic field, the quantum CME oscillations dominate over the SdH ones; these CME oscillations have a phase of $-\pi/4$. On the other hand, in weak fields the SdH oscillations are dominant, with the phase of $\pi/4$. 

\vskip0.3cm

To summarize, we have demonstrated that the non-linear relation between the density of chiral charge and the chiral chemical potential induces a new type of quantum oscillations in longitudinal magnetoconductivity of Dirac and Weyl (semi)metals. In strong magnetic fields and in materials that approximately preserve chirality (when the chirality flipping time is much longer than the scattering time), these new quantum oscillations dominate over the SdH ones. The  phase of these quantum CME oscillations differs from the SdH oscillations by $\pi/2$ which makes it possible to 
isolate them in experiment.

\vskip0.3cm

We thank Qiang Li for useful and stimulating discussions. This work was supported in part by the U.S.
Department of Energy under Contracts No. DE-FG- 88ER40388 and DE-AC02-98CH10886, and by the LDRD 16-004 at Brookhaven National Laboratory.

\bibliography{references}

\end{document}